\documentclass[aps,showpacs,pra,twocolumn]{revtex4}
\usepackage{epsfig,amssymb,amscd}

\newcommand{\ket}[1]{|#1\rangle}

\begin{document}
\bibliographystyle{prsty}
\draft

\title{A quantum computer using a trapped-ion spin molecule and microwave radiation}
\author{D. Mc Hugh, J. Twamley\footnote[2]{Email: Jason.Twamley@may.ie}}
\affiliation{Department of Mathematical Physics,\\
National University of Ireland Maynooth,\\ Maynooth, Co. Kildare, Ireland}
\received{\today}

\begin{abstract}
We propose a new design for a quantum information processor where qubits are encoded into Hyperfine states of ions held in 
a linear array of individually tailored linear microtraps and sitting in a 
spatially varying magnetic field. The magnetic field gradient introduces spatially dependent qubit transition frequencies and 
a type of spin-spin interaction between qubits. Single and multi-qubit  manipulation is achieved via resonant microwave pulses as in liquid-NMR quantum computation while the qubit readout and reset is achieved through trapped-ion fluorescence shelving techniques.
By adjusting the microtrap configurations we can tailor, in hardware, the qubit resonance frequencies and coupling strengths.   
We show the system possesses a side-band transition structure which does not scale with the size of the processor allowing scalable frequency
discrimination between qubits. By using large magnetic field gradients, one can reset individual qubits in the ion chain via frequency selective {\it optical} pulses to implement quantum error correction thus avoiding the need for many tightly focused laser beams. \\
\end{abstract}
\vskip 0.1cm
\pacs{03.67.-a}
\maketitle

A quantum computer requires well-characterized quantum bits 
(qubits), sufficient control to perform a universal set of quantum gates, 
long decoherence times compared to gate times and a means of performing qubit 
readout \cite{DiVincenzo}. 
To be of practical use, any design must scale to hundreds of 
qubits without enormous technical overhead.
Current ion trap and nuclear magnetic resonance (NMR), implementations 
are capable of satisfying most of these criteria to varying degrees.
Liquid-state NMR quantum computing has demonstrated very precise control in 
the manipulation 
of qubits with microwave and radio frequency pulses and the execution of a 
number of quantum algorithms.
The main drawback of Liquid-NMR quantum computing is the extreme
difficulty, due to tiny spin polarization and weak measurements, in scaling up 
to a large number of qubits.
Ion trap quantum computing has recently shown progress
in implementing a quantum algorithm \cite{blatt_deutsch}, and a controlled 
interaction between two qubits \cite{blatt_zollergate}. Here the initialization
and readout of qubits is particularly good. However, there are 
barriers to be overcome such as decoherence due to ion heating, very precise 
laser focussing and stability problems of operating at optical frequencies. 
These difficulties may hinder the ability to scale optically-addressed 
ion-trap technologies to hundreds of ions although a number of potentially scalable designs involving 
moving and stationary optically manipulated trapped ions have been proposed \cite{Jonathan}. In this article we describe 
a new design for a potentially highly scalable quantum information processor 
which combines the trapped-ion and NMR technologies in a manner that
retains the advantages of both. \\

We assume that two Hyperfine levels in each ion serve as the qubit. 
As demonstrated by Wunderlich {\it et. al.} \cite{Mintert, Wunderlich,Balzer}, 
one can induce a coupling between the qubits  through the application of 
a magnetic field gradient  along a string of ions. This, together with the Coloumb force, 
couples the qubits together in a way analogous to the spin-spin coupling observed between nuclear 
spins in a molecule.
Obviously then, similar techniques to those used in NMR, and the related field of electron spin resonance (ESR),
can be utilised to realize two and multi-qubit quantum gates. 
The work presented by Wunderlich {\it et. al.}, though possessing obvious benefits for the precise manipulation of quantum information, however displays a number of serious obstacles with respect to scalability. In that design 
we will find that the spin-spin coupling varies in strength throughout the ion chain thus leading to varying quantum gate durations between qubits depending on their location in the ion chain. This, itself poses a serious difficulty in that a scalable device will require device-wide parallel quantum information processing and quantum error correction and this will need careful synchronisation of logical operations throughout the device. More problematic is the existence in the Wunderlich design of an upper limit to the number of qubits which can be cleanly frequency differentiated for a given size of magnetic field gradient. Finally, the technical difficulty and potential added source of decoherence introduced by the focusing of individual laser beams on each ion in the chain for \emph{individual} qubit initialization, readout and reset may also be problematic for scalability \cite{Balzer}. Here we show that  
by considering a design where the ions are stored in a linear array of individually tailored linear microtraps \cite{Eliptical}, we can surmount all of the above mentioned difficulties and achieve significant gains in scalability.
In particular we find that the resulting system is, as before, analogous to an $N$ atom molecule with spin-spin couplings, but now these couplings can be tailored very precisely, and quite robustly,  through altering the individual microtrap parameters.
The new design is also exactly analogous to a true ion crystal where each ion is harmonically bound to a periodic spatial lattice. The resulting qubit resonance frequencies together with the vibrational side-transitions form a band structure (as in condensed matter systems), where a band's extent now only depends on the inter-ion spacing and end trap strength and does not grow with the number of ions in the chain.
Finally, as previously shown by Wunderlich \cite{Wunderlich}, the magnetic field gradient  allows the qubits to be individually frequency-addressed in the microwave via Zeeman splitting of the Hyperfine structure. We further show that for large magnetic field gradients one can engineer for {\em frequency-addressed  optical resetting} of an individual Hyperfine qubit via Zeeman splitting of the {\em optical} transitions. This capability allows for scalable quantum error correction and  relaxes the need for focusing of individual laser beams on each ion through the use of frequency multiplexed laser sources.  Previous difficulties predicted in liquid NMR quantum information processing related to the 
saturation of the available RF-bandwidth with a moderate number of frequency differentiated qubits is no longer an issue here as the qubit manipulation in the MW, (or readout in the optical), has available radiation sources   which are tunable over many MHz (GHz).
Further, our design uses the highly developed and relatively widespread technology of microwave pulse synthesis to precisely manipulate the qubits. This avoids the need to develop and maintain ultra-stable narrow-band laser systems as used in 
 in \cite{blatt_deutsch,blatt_zollergate}, although high frequency optical systems will still be required for the qubit initialization, reset and readout processes.  
Our model differs fundamentally from other quantum computer designs based on trapped ions. In \cite{CZ95}, a two-qubit gate is realised
through the exchange of a phonon through collective vibrational motion of the ions while in \cite{CZ405}, the ions are also stored in
an array of microtraps and qubit gates are realised through a state-dependent displacement of the motional wave-packet of the ions. 
In the scheme of Wunderlich and in our scheme, the spin-spin coupling is achieved via virtual excitation of the ion's motion and thus gate operations should be more robust against decoherence arising from the motion of the ions as in the schemes presented in \cite{Molmer,Milburn,CirZolRip}.  \\

We first explain qualitatively how the model operates and follow this with a
quantitative analysis. We consider $N$ ions, with each ion occupying a
separate harmonic oscillator
potential well, arranged in a linear array. We assume we can independently 
fix (1), the strength of the each potential well and, (2) the separation 
between each well.
Each ion stores a qubit in two Hyperfine states and taken together constitutes a quantum 
register. The 
initialization of the qubits is performed as usual via fluorescence shelving and repumping \cite{Sasura02}.
A magnetic field gradient is applied along the ion string and results in frequency differentiated qubits in the string due to the Zeeman shifting of each ion's Hyperfine levels. 
Single-qubit
operations on a given ion are performed by illuminating the ion string with  
a pulse of microwave
radiation of the appropriate frequency. 
The spin-spin  coupling between the ions is achieved through a combination of the 
Coloumb force and the qubit dependent Zeeman energy which is spatially 
modulated by the magnetic field gradient \cite{Wunderlich}.
The resulting spin-spin coupling is thus ``on" all the time and 
gates are achieved via a combination of single-qubit operations and free evolution as in liquid-NMR quantum computation. 
Thus one can use the considerable knowledge of 
NMR refocusing and averaging techniques \cite{Beth}, to tailor the system 
Hamiltonian and perform quantum information processing. 
However, quantum computing using liquid state NMR and an ion spin-molecule sitting in a single trapping potential as proposed by Wunderlich, 
differ in that the latter allows one to control (although in a very rough manner), {\em  in 
hardware}, the inter-qubit coupling strengths. 
For the latter the resulting inter-qubit couplings vary in strength throughout the ion-chain no matter how one alters the single trap parameters (see Fig. \ref{Fig2}). This lack of homogeneity represents a significant barrier for the scalability of \cite{Wunderlich}.
In addition, using a single linear trap to contain the ion string yields limits on the number of ions in the string that can be cleanly frequency differentiated for a given size of magnetic field gradient.  
By introducing linear microtraps and individually tailoring their strengths and locations we can surmount both of the above problems to yield a design with high scalability. We can precisely engineer the inter-qubit couplings to be homogenous throughout the chain while maintaining uniform separation between the ions and consequently, well differentiated qubit resonant frequencies irrespective of the size of the device.  
Having an homogeneous system enormously boosts the scalability of device as the alternative would imply that all gate operations (pulse sequences), would be highly dependent on the location  of the ions in the chain which are involved in the gate.  In addition by varying the ion-separation and microtrap strengths we also have some control over the relative sizes of non-nearest-neighbour to nearest-neighbour spin-spin coupling strengths. 
In \cite{Wunderlich}, and in our model, the ions in the string are spatially separated on length scales varying from $2-10 \mu$m, and although the individual focusing of lasers on each ion is possible, such a readout scheme may yield too much decoherence due to laser pointing fluctuations, amplitude noise etc. 
Also in \cite{Wunderlich}, one can readout the entire qubit register at once at the completion of a quantum algorithm. However to execute quantum error correction, intermediate measurements are usually desired. However, as shown in \cite{nielsen}, quantum error correction only requires the capability of supplying fresh ancilla qubits, or in the present case, the resetting of any individual qubit to a preset quantum state. We show that in the case of large constant magnetic field gradients, 
the resulting spatial dependence of the energies  
of both the qubit Hyperfine levels {\em and} the optically-excited ``readout'' level 
allows one to achieved optical frequency differentiated reset through optical pumping or through engineered decoherence processes, with little disturbance of neighboring qubits. This yields another boost for the scalability of the resulting design. \\

We now look at the model more quantitatively. Consider $N$ ions each of mass
$m$ confined in $N$ individual harmonic potential wells. The $i^{th}$ linear trap
has frequency
$\omega_i$ and is located at position $k_i$ along the $x$-axis. 
A spatially varying magnetic field, $\vec{B}\equiv (B_0+b x)\hat{z}$,  
is applied across the line of ions. The resulting Hamiltonian, to second order in the ion's vibrational motion, is given by \cite{Wunderlich}, 

\begin{equation}
 H=\frac{\hbar}{2}\sum^N_{n=1}\omega_n(x_{0,n})\sigma_{z,n}+\sum^N_{n=1}\hbar\nu_n a^{\dagger}_na_n-\frac{\hbar}{2}\sum_{n<m}J_{nm}\sigma_{z,n}\sigma_{z,m}\;\;.
\label{bigh}
\end{equation}

The first term represents the electronic Hamiltonian of the qubits, now with 
separated qubit resonances $\omega_n(x_{0,n})$, and where $x_{0,n}$ is the equilibrium location of the $n$'th
ion in the string sitting in the trapping potential. The second term describes the 
collective quantized
vibrational motion of the ions. Even though the ions are now in individual linear 
traps, this term describes the quadratic interactions between the ions and is essentially identical 
to the case of a  single harmonic trap. The last term expresses the pairwise 
coupling between qubits analogous to the well-known spin-spin coupling in
molecules used for NMR quantum computing. 
In 
\cite{Wunderlich}, it is then proposed that this last term  
can be used to implement quantum gates.\\

One can show that the spin-spin coupling between ions $n$ and $m$ can be expressed in the relatively simple form:

\begin{eqnarray}
J_{nm} & = & \frac{\hbar}{4\pi m}\sum^N_{j=1}\frac{1}{\nu_j^2}\left. \frac{\partial\omega_n}{\partial x_n}\right|_{x_{0,n}}\left. \frac{\partial\omega_m}{\partial x_m}\right|_{x_{0,m}} D_{nj}D_{mj}\\
&=&\frac{\hbar}{4\pi}\left.\frac{\partial \omega}{\partial x}\right|_n\left.\frac{\partial \omega}{\partial x}\right|_m(A^{-1})_{n\,m}\;\;,\label{bigJ}
\end{eqnarray}

\noindent
where $\left.\frac{\partial \omega}{\partial x}\right|_k$ is the gradient of 
the qubit transition resonant frequency for ion $k$, at equilibrium location $x_{0,k}$; 
$A$ is the Hessian of the potential 
in which the ions sit; $D$ is the unitary transformation matrix that diagonalises $A$ and $m\nu_j^2$ are the eigenvalues of $A$. 

In the high-field or Paschen-Bach limit, ($B_0 \sim 1$T), 
the frequency gradients are independent of $x$ and thus $n$, and so $J\propto A^{-1}$.
To achieve uniform off-diagonal values for $J\sim A^{-1}$, i.e. $J_{ij}\sim J_{|i-j|}$, we now tailor the values of the Hessian $A$.
Firstly the off-diagonal elements of $A$ are functions only of the inter-ion
spatial separations. We can fix the positions of the individual microtraps, $k_i$, to achieve
a uniform ion separation, $h$, thus enormously simplifying the structure of $A$. 
The diagonal elements of $A$ are functions of both the inter-ion separation, $h$, and the individual trap strengths.
Letting $g_i\equiv m\omega_i^2$, be a measure of the $i$'th trap strength, 
$\epsilon_1=(e^2/4\pi \epsilon_0 h)/(\frac{1}{2} g_1 h^2)$ \cite{CZ405}, and assuming only nearest neighbor coupling, one can analytically show that for uniform off-diagonal $J$'s one must have 
\begin{equation}
g_1=g_N\;\;,\;\; g_i=\frac{g_1}{1+\epsilon_1}\;\;, \;\;1< i < N\;\;,
\label{strengths}
\end{equation}
with all interior traps being weaker in strength than the end traps. Resorting to numerics to include non-nearest neighbor terms in the Hessian, one can determine the interior microtrap strengths for any given length of ion string with a uniform spatial ion separation. As one might expect both $k_i$ and $g_i$ are symmetrically valued about the mid-point of the ion chain. 
As an example, for Yb${}^+$ ions, where the qubit is encoded in the Hyperfine states $\ket{0} \equiv \ket{6\ S_{\frac{1}{2}},\ M_I=\frac{1}{2},\ M_J=-\frac{1}{2}},\;\;
\ket{1}\equiv \ket{6\ S_{\frac{1}{2}},\ M_I=\frac{1}{2} M_J=\frac{1}{2}}$, of each ion 
we can obtain the microtrap strengths in Fig. \ref{Fig3}. Once suitable values for $g_i$ are obtained, the resulting microtrap positions $k_i$, which yield the specified uniform ion separation, can be separately computed numerically. As expected, the microtrap separation decreases slightly towards the edges of the ion string to compensate for the larger outwards Coloumb forces experienced by ions in these regions.  
From Fig. \ref{Fig2}, the resulting improvement in the uniformity of the $J$-couplings throughout the ion string is quite striking. In addition however, we have also succeeded in significantly reducing the non-nearest neighbor $J$-couplings, thus reducing these potential sources of error in the performance of nearest neighbor quantum gates. Further, by varying both the inter-ion separation $h$, and end-trap strength $g_1$, we can control the degree of suppression of non-nearest neighbor couplings. Thus, through tailoring the microtrap configuration we can achieve uniform off-diagonal coupling strengths throughout the ion-string. Other types of $J$-couplings may also be tailored for.\\

Of course it is important to estimate the degree of intolerance of the design due to 
misalignments in the trap strengths and positions. 
In numerical simulations, fluctuations of the ion positions
have minimal effect on the uniformity of the $J$-couplings. Given 10 ions in a trap, 
with $\nu_1=1$MHz,  $b=1000$T/m and 
$h=$10$\mu$m, setting $x_{0,n}=x_{0,n}^U+\delta x_n$, where $x_{0,n}^U$ are the microtrap locations giving a uniform $J$-coupling, and $\delta x_n$  randomly distributed deviations within a range $\pm 0.1\mu{\rm m}$,  
the nearest-neighbour $J$-couplings vary by 
approximately 1\% on average. Considering the spatial extent of the lowest 
axial vibrational mode of each ion is $\sim$2nm, the $J$-couplings maintain
their uniformity when heating occurs. For variations of the individual linear microtrap strengths, $\nu_i=\nu_i^U+\delta\nu_i$, with $\delta\nu_i$ 
randomly distributed deviations within a range  $\pm 10{\rm kHz}$ ($\sim \pm 1\%$ of $ \nu_1$), results in 1\% variations of the nearest-neighbour $J-$couplings. Further, one can employ so-called robust pulse sequences to overcome small inhomogeneities in the coupling strengths throughout the chain \cite{jones}.

We now take some time to outline the complete scalability of the qubit frequency discrimination within our design. As noted in \cite{Wunderlich,Balzer}, the vibrational energy of the highest collective motional mode of $N$ ions in a single harmonic trap subject to a magnetic gradient of size $\partial B/\partial x=b$, increases linearly with $N$, i.e. $\nu_{max}\sim \nu^* N$. For clean frequency discrimination between two neighboring qubit resonant frequencies, $\omega_i$ and $\omega_{i+1}$, one requires, $|\omega_{i+1}-\omega_i|\propto \partial B/\partial x >2\nu_{max}\sim 2\nu^*N$. Thus for  a given magnetic field gradient $b$, there is a maximum number of ions, $N^*$, which can be cleanly discriminated. In our model, however, the vibrational frequencies of the collective motion of $N$ ions, held in microtraps tailored for uniform spatial separation and qubit coupling strengths $J$,  asymptotes  to a given value \mbox{$\nu_{max}(N,h,g_1)\rightarrow \nu_{max}(h,g_1)$}, for large $N$. Empirically, defining  $\Delta \nu(N,h,g_1)\equiv \nu_{max}(h,g_1)-\nu_{max}(N,h,g_1)$, we find $\Delta\nu(N,h,g_1)\sim \Delta\nu(N_j,h,g_1)\times (N/N_j)^{-1.78}$. Empirically, we also can find the scaling of this convergence with ion separation $h$, to be $\Delta\nu(N,\tilde{h}, g_1)\sim \Delta\nu(N,h,g_1)\times (\tilde{h}/h)^{-3}$. \\

It is necessary finally, to discuss in more detail how it is proposed that 
the qubits can be individually manipulated, read out and reset. First, we will talk
about manipulation. Since we are proposing to using two Hyperfine levels 
for the qubit, long wavelength radiation is involved and so we will have to
manipulate the qubits via  frequency selection rather than by attempting to focus laser light on the 
ions individually. As an example, we again identify the 
$\ket{S_{\frac{1}{2}}, M_I=\frac{1}{2}, M_J=-\frac{1}{2}}$ and 
$\ket{S_{\frac{1}{2}}, M_I=\frac{1}{2}, M_J=\frac{1}{2}}$, states of 
${}^{171}$Yb$^+$ as the qubit 
$\ket{0}$ and $\ket{1}$. The Zeeman splitting of the Hyperfine levels lies in the Paschen-Back region when $\frac{g_J\mu_BB(z)}{E_{HFS}} > 1$, 
where $E_{HFS}$ is the Hyperfine splitting constant, $g_J$ the g-factor and 
$B(z)$ the strength of the magnetic field. For Yb${}^+$, this gives $B>1$T. If we are in this
region then the difference between the manipulation frequencies of neighbouring
qubits is given by $\Delta\nu_{manipulate} =\frac{g_J\mu_Bb\delta z}{h}$,
where $b=\frac{\partial B(z)}{\partial z}$ and $\delta z$ is the separation 
between the ions. Choosing a field gradient of $1000$ T/m and a typical ion 
separation distance of $10\mu$m we can realise a neighbouring qubit frequency difference
of $280$MHz, irrespective of the length of the ion-string. 
This separation is easily differentiated by current ESR Microwave pulse spectrometers.

The final readout of the state of the qubits can be performed, as usual, by spatially resolved resonance fluorescence. 
However, scalability will require the capability to perform quantum error correction. This capability will require individual ion readout and/or reset.
However, as the spatial separation between ions is $\sim\mu$m, readout/reset of individual ions
through individual laser focusing may be difficult. 
Instead,  one can use frequency addressing of the optical transitions to achieve individual ion readout/reset. 
For $B>1T$, both readout levels, 
$|6S_{\frac{1}{2}},M_I=\frac{1}{2},M_J=\frac{1}{2}\rangle$ and
$|6P_{\frac{1}{2}},M_I=\frac{1}{2},M_J=-\frac{1}{2}\rangle$,
are in the Paschen-Back region and the resulting frequency separation between the optical transitions of neighbouring 
qubits is $\Delta\nu_{readout} =\frac{4}{3}\frac{\mu_Bb\delta z}{h}$. This 
gives an optical frequency difference of $187$MHz for the above situation which is far in excess of the excited state lifetime $\sim 10$MHz, and thus cleanly discriminated between neighboring ions.
Quantum error correction  typically involves 
the measurement of an individual qubit state but to use fluorescence detection in a manner which leaves the spectator qubits undisturbed requires the additional correction of considerable AC Stark shifts
suffered by neighboring ions during the illumination and acquisition of $\sim 10^{4}$ fluorescent photons from the target ion. Techniques
to perform this correction have been demonstrated \cite{Haffner}, but such invasive detection may not be necessary as only  the capability to perform a {\em qubit specific reset}, are required for quantum error correction \cite{nielsen}. Resetting the Hyperfine qubit to a known quantum state should only require the scattering of far fewer fluorescent photons (\# of photons $\sim M=10^0-10^1$), via optical pumping and the resulting phase acquired by a neighboring ion off-resonant by $L\sim 19$, radiative linewidths, will be $\phi\sim M/L\sim 1/5$ radians. The correction of such phases are well within the capabilities of the  techniques demonstrated in \cite{Haffner}. Alternatively, a technique to reset an individual  Hyperfine qubit by simultaneously resonantly irradiating the qubit and optical transitions with strong MW and extremely weak optical radiation at 369nm (with a laser power of nW), has previously been demonstrated in \cite{Balzer2002}. This process of {\em engineered decoherence}, may also be used as a qubit reset.

Recent applications of narrowband ultraviolet laser diodes diode lasers to Ytterbium trapping might be adapted for
the  frequency selective optical addressing \cite{UVLD}. By using electro-optic modulators and external cavity tuning elements, tuning ranges of several GHz are currently possible at these wavelengths. Through illumination of the entire ion chain by  a frequency multiplexed laser source one can achieve individual qubit readout and reset in a highly scalable manner. Achieving in-trap magnetic field gradients of
$~1000$T/m over several micron can be achieved through methods similar to those presented in \cite{gradients}. Also, engineering linear ion traps on the micron scale is possible as described in \cite{microtraps}.
However, by decreasing $h$, we increase the qubit coupling strengths while decreasing the frequency separation between the qubit manipulation and readout frequencies. 
As a final example, in Table 1, we have computed the parameters for ten Yb${}^+$ ions, with $B_0=1$T, $b=1000$T/m, $\nu_1=1$MHz, and $h=10\mu$m, held in individually tailored microtraps for uniform separation and $J-$couplings. Here the uniform nearest neighbour 
coupling strength, $J_{i,i+1}\sim 850$Hz. The tabulated parameters are within current microwave source capabilities and clearly indicate  scalable frequency discrimination for the manipulation,  readout and reset of  neighboring qubits. \\

In conclusion, we have presented here a new design for a trapped-ion spin-molecule quantum information processor which we feel should be highly scalable. Our design  surmounts the difficulties in scalability faced by the scheme of Wunderlich {\it et.al.} By storing the ions in individually tailored linear microtraps we can engineer homogenous coupling strengths throughout the chain while simultaneously achieving a constant frequency separation between neighboring qubits, irrespective of the length of the ion chain. This enormously simplifies the microwave pulse sequences required to manipulate the quantum information in the ion chain. At the same time we have removed the limit on ion numbers which can now be cleanly frequency discriminated in a fully scalable manner. Finally, we can utilise the spatial dependence of the energy level structure of the optical transitions to achieve qubit frequency selective reset and initialisation, tasks which are essential for the ultimate fault tolerant operation of a scalable device. \\

{\bf Acknowledgement}\\
D. McHugh kindly acknowledges support from Enterprise-Ireland Basic Research Grant SC/1999/080.\\

\newpage
\begin{table}[H]
\begin{center}
\begin{tabular}{|l|l||}
\hline\hline
Mean Qubit Resonance Frequency (GHz) & 34.3 \\\hline
Neighboring Qubit Resonance Frequency Separation (MHz) & 280 \\ \hline
Motional Sideband Extent  (MHz) & 3\\ \hline
Neighboring Qubit Optical Readout Frequency Separation (MHz) & 187\\ 
\hline\hline
\end{tabular}
\caption{Relevant parameters for ten Yb${}^+$ ions, with $B_0=1$T, $b=1000$T/m, $\nu_1=1$MHz, and $h=10\mu$m, held in individually tailored microtraps for uniform separation and $J-$couplings.}
\label{table1}
\end{center}
\end{table}

\begin{figure}[p]
\begin{center}
\setlength{\unitlength}{1cm}
\begin{picture}(6,10)
\put(-3.0,4.5){\Large (a)}
\put(4.4,4.5){\Large (b)}
\end{picture}
\end{center}
\caption{Qubit coupling strengths, $J_{ij}/2\pi$, in Hz, for a ten Yb${}^+$ ion string in a magnetic field $B(x)=B_0+b x$, with $B_0=1$T,  $b=1000$ T/m, where (a) the ion string is held in a single harmonic trap of strength $1$MHz, and (b) where each ion is held at a uniform spatial separation of $10\mu$m, via individually tailored microtraps with an end trap strength of $1$MHz.}
\label{Fig2}
\end{figure}

\begin{figure}[p]
\begin{center}
\setlength{\unitlength}{1cm}
\begin{picture}(6,10)
\end{picture}
\end{center}
\caption{Microtrap frequencies, $\omega_j$,  as a function of ion-chain length, $N$, for a uniform ion spacing of $10\mu m$, end-trap frequency of 1MHz and magnetic field gradient of $1000$T/m. Microtrap frequencies are symmetrically distributed about the geometric center of the ion-chain, curve \#k refers to the microtraps $j=k+1, N-k$, in a chain of length $N$. }\label{Fig3}
\end{figure}


\begin{thebibliography}{17}
\bibitem{DiVincenzo} D. P. DiVincenzo, {\it Fortschr. Phys.} {\bf 48}, 771-783 (2000).
\bibitem{blatt_deutsch}S. Gulde, M. Riebe, G. Lancaster, C. Becher, J. Eschner, H. H\"{a}ffner, F. Schmidt-Kaler, I. L. Chuang \& R. Blatt, {\it Nature} {\bf 421}, 48-50 (2003). 
\bibitem{blatt_zollergate}F. Schmidt-Kaler, H. H\"{a}ffner, M. Riebe, S. Gulde, G. P. T. Lancaster, T. Deuschle, C. Becher, C.F. Roos, J. Eschner \&  R. Blatt, {\it Nature} {\bf 422}, 408 (2003);  D. Leibfried, B. DeMarco, V. Meyer, D. Lucas, M. Barrett, J. Britton, W. M. Itano, B. Jelenkovi\'{c}, C. Langer, T. Rosenband \&  D. J. Wineland, {\it Nature} {\bf  422}, 412 (2003).
\bibitem{Jonathan}D. Jonathan, M. B. Plenio, \& P. L. Knight, Phys. Rev. A {\bf 62}, 042307 (2000); D. Kielpinski, C. Monroe, \& D. J. Wineland, Nature {\bf 417}, 709 (2002); C. Monroe, Nature {\bf 416}, 238 (2002); L. M. Duan, B. B. Blinov, D. L. Moehring \& C. Monroe, quant-ph/0401020; J. J. Garcia-Ripoll, P. Zoller \&J. I. Cirac, quant-ph/0306006; L. M. Duan, quant-ph/0401185.
\bibitem{Mintert}F. Mintert \& C. Wunderlich, {\it Phys. Rev. Lett.} {\bf 87}, 257904 (2001).
\bibitem{Wunderlich}C. Wunderlich, in {\it Laser Physics at the Limit}, edited by H. Figger, D. Meschede \& C. Zimmermann (Springer-Verlag, Berlin, 2001), p. 261-271. 
\bibitem{Balzer}C. Wunderlich, C. Blazer, T. Hannemann, F. Mintert, W. Neuhauser, D. Rei\ss \& P.E. Toschek, {\it J. Phys. B} {\bf 36} 1063-1072 (2003).
\bibitem{Eliptical}R. G. DeVoe, {\it Phys. Rev. A} {\bf  58}, 910-914 (1998).
\bibitem{nielsen}M. A. Nielsen \& I. L. Chuang, Quantum computation and quantum information (Cambridge University Press, 2000), p. 439.
\bibitem{CZ95}J. I. Cirac \&  P. Zoller, {\it Phys. Rev. Lett.} {\bf 74}, 4091-4094 (1995).
\bibitem{CZ405} J. I. Cirac \& P. Zoller, {\it Nature}  {\bf 404}, 579–581 (2000).
\bibitem{Sasura02} M. Sasura \& V. Bu\v{z}ek, {\it J. Mod. Opt.} {\bf 49}, 1593-1647 (2003).
\bibitem{Molmer} A. S\"{o}renson \& K. M\"{o}lmer, {\it Phys. Rev. Lett.} {\bf 82}, 1971 (1999).
\bibitem{Milburn} G. Milburn, S. Schneider \& D.F.V. James {\it Fortschr. Phys.} {\bf 48}, 801-810 (2000).
\bibitem{CirZolRip} J.J. Garcia-Ripoll, P. Zoller \& J.I. Cirac {\it Phys. Rev. Lett.} {\bf 91}, 157901 (2003).
\bibitem{Beth}R. R. Ernst, G. Bodenhausen \&  A. Wokaun {\it Principles of Nuclear Magnetic Resonance in One and Two Dimensions}, (Oxford University Press, Oxford, UK, 1990); P. Wocjan, M. Rotteler, D. Janzing \&  T. Beth, {\it Phys. Rev. A} {\bf 65}, 042309 (2002). 
\bibitem{jones}J. A. Jones, Phys. Rev. A {\bf 67}, 012317 (2003).
\bibitem{Haffner}H. H\"{a}ffner,  S. Gulde, M. Riebe, G. Lancaster, C. Becher, J. Eschner, F. Schmidt-Kaler, \& R. Blatt, Phys. Rev. Lett. {\bf 90}, 143602 (2003).
\bibitem{Balzer2002}C. Balzer, T. Hannemann, D. Rei§, W. Neuhauser, P. E. Toschek, \&
C. Wunderlich, Laser Physics, {\bf 12}, 729 (2002).
\bibitem{UVLD}C. Y. Park, T. H. Yoon, J. I. Kim, K. Y. Yeom \&  E. B. Kim {\it Optics Lett.} {\bf 28}, 245 (2003).
\bibitem{gradients}R. Folman, P. KrŸger, D. Cassettari, B. Hessmo, T. Maier \&   J. Schmiedmayer, {\it  Phys. Rev. Lett.} {\bf 84}, 4749 (2000);  D. Suter \&  K. Lim {\it Phys. Rev. A} {\bf 65}, 052309 (2002).
\bibitem{microtraps}M.J. Madsen, W.K. Hensinger, D. Stick, J.A. Rabchuk, C. Monroe, {\it e-print} {\bf quant-ph/0401047}, (2004).
\end{thebibliography}
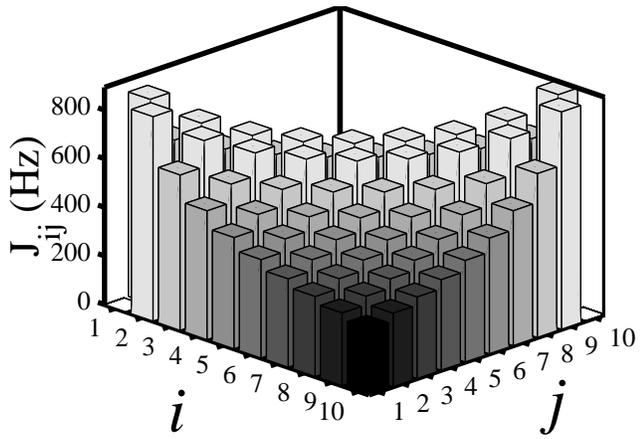
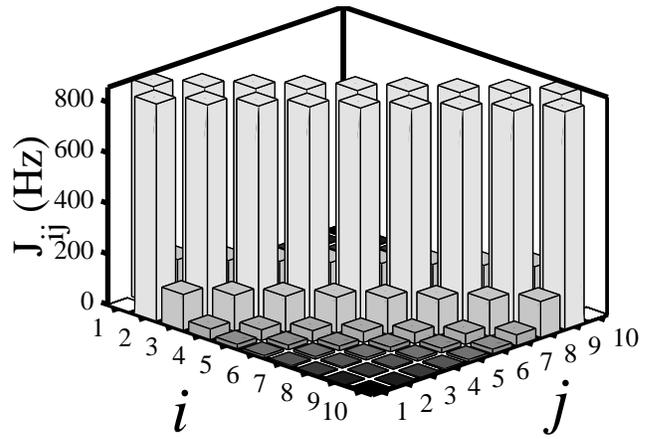
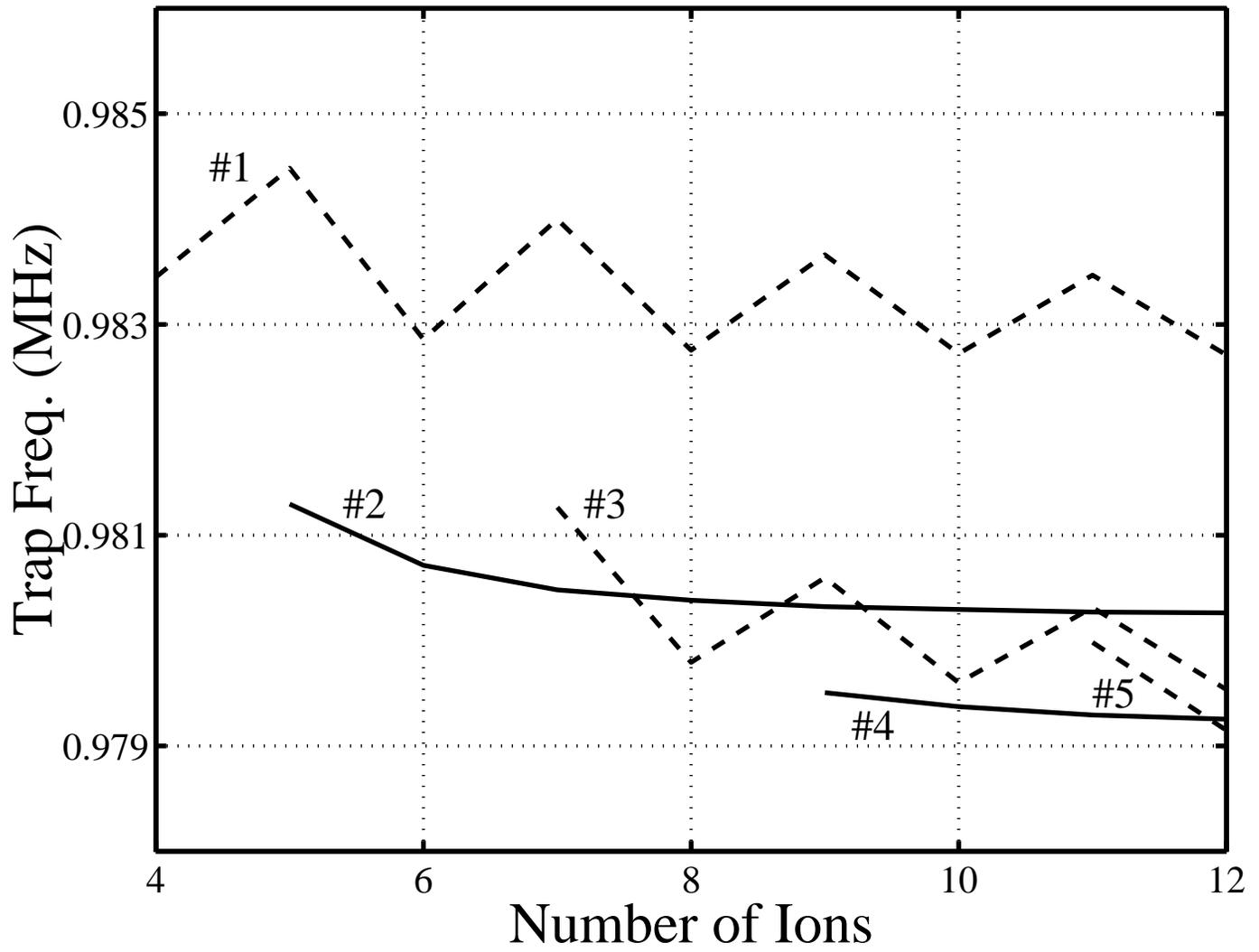
\end{document}